# Transient chirp reconstruction of ultrafast electron beam via tightly focused chirped laser pulse

Zhijun Zhang[1, 2], Shiyi Zhou[2], Jiansheng Liu[2]

[1]Yongjiang Laboratory, Ningbo 315202, China

[2]Department of Physics, Shanghai Normal University, Shanghai 200234, China

Controlling the phase space of particle beams is essential for the generation of high-quality, ultrashort electron beams in plasma-based accelerators. Accurately diagnosing the transient energy chirp, which evolves rapidly at the onset of acceleration, presents a significant challenge. This paper introduces a novel method for reconstructing the transient energy chirp of ultrashort electron beams by employing tightly focused and chirped laser pulses. We investigate the conditions that enhance the modulation of electron-beam divergence and illustrate the reconstruction of transient chirp based on the intrinsic phase correlation of the modulated divergence as projected onto specific phase space coordinates. Additionally, we estimate the temporal delay between the laser and electron beam by applying a Fourier transform to the reconstructed divergence modulation in the frequency domain. This approach holds promise for optimizing accelerator performance and facilitates the probing of timing jitter in ultrafast electron diffraction with attosecond-level precision.

## 1. Introduction

Pump-probe techniques utilizing ultrashort X-rays or electrons as probe pulses have enabled the measurement of ultrafast phenomena in atomic and molecular structures [1, 2]. The temporal resolution of these techniques is mainly determined by the pulse duration and arrival time jitter of the probe pulse. In recent years, significant progress has been made in generating femtosecond pulses from X-ray free-electron laser facilities [3-9], ultrafast electron diffraction (UED) facilities [10-13] and plasma accelerators (PAs) [14-16]. Electron beams (e-beams) from plasma accelerators typically have durations of a few femtoseconds, owing to the inherently microscopic plasma wake (~10 μm for plasma with density of $10^{19}$ $cm^{-3}$). Compact table-top laser wakefield accelerators generating high brightness GeV e-beams with energy spread at the few per-mille level have been developed, [17, 18], and their feasibility of free electron lasing has been demonstrated [19]. As a result, the PA-based ultra-short pump-probe technique is becoming feasible and promising.

To enhance the capabilities of PAs, extensive efforts have been devoted to improving the quality of e beams. These efforts include compression of the energy spread [20-25] and manipulation of the duration towards attosecond sources [26-28]. Achieving precise control over the e-beam phase space distribution, which evolves rapidly in the extreme wakefield gradients, is crucial for these schemes.

However, a technique for accurately measuring the transient chirp profile of the e-beam has not yet been established. While transverse RF deflectors have enabled chirp reconstruction of femtosecond e-beams [10, 29, 30], they require a low energy spread of the e-beam at the per-mille level to prevent phase space rotation during the long-distance interaction (typically 1 m for the RF deflector). Additionally, for measuring the timing jitter between the pump pulse and the probe beam, the terahertz streaking technique has been employed for femtosecond-resolved measurements but necessitates the use of a complex terahertz source [10, 31]. Therefore, there is a need for a more convenient technique that offers higher temporal resolution to facilitate the reconstruction of transient phase space and probing of timing jitter in UED experiments.

Here, we present a novel method to reconstruct the transient energy chirp of an electron beam utilizing tightly focused and chirped laser pulses. By tightly focusing the laser pulse and introducing a suitable propagation crossing angle (θ) between the electron beam and the laser pulse, the interaction between the two is confined to a short time frame. Consequently, the electron beam undergoes longitudinal modulation due to the interaction with the linearly polarized laser field. To determine the chirp profile of the electron beam, we measure the projection of this modulation along the momentum coordinate using an electron spectrometer. By correlating the projected modulation with the laser field distribution, the chirp profile of the electron beam can be accurately reconstructed. Moreover, by applying a chirped laser pulse, it becomes possible to determine the delay between the laser and electron beam, making it suitable for probing timing jitter in UED experiments. The temporal resolution of this method is determined by the laser period, and with the utilization of high-efficiency laser harmonic techniques[35], tunable and high-resolution measurements down to the attosecond scale can be achieved. This robust method is easy to implement and can be applied for both optimizing PAs and probing timing jitter in UED experiments, potentially reaching temporal resolutions at the attosecond level.

## 2. The laser electron interplay

Based on the Lawson–Woodward theorem [36], it is known that when electrons interact with a temporal symmetrical pulse over an infinite interaction region, there is no net energy gain. To overcome this limitation, various schemes have been proposed. For example, the use of a chirped laser pulse [37] or the confinement of the interaction region [38] can generate asymmetric fields,, and it has been demonstrated that introducing an obliquely incident laser to limit the interaction length can modulate the momentum of electrons [39], which is widely employed in free-electron laser facilities for electron pre-bunching [40]. In our proposed scheme, depicted in Fig. 1, we introduce a crossing angle ($\theta$) between the laser and the electron beam. This leads to the imprinting of the electron beam along the ζ direction with longitudinally dispersed transverse momentum gain. Here, $\zeta$ represents the retarded time ($\zeta$=$z$-$ct$), where c is the speed of light in vacuum. The dispersed transverse momentum gain manifests as a divergence modulation characterized by $\alpha(\zeta)$. The modulated electron beam is subsequently measured using an electron spectrometer, enabling the detection of the projected modulation $\alpha(p_z)$. The energy chirp, represented by the $p_z - \zeta$ correlation, can be estimated by analyzing the correlation between $\alpha(p_z)$ and $\alpha(\zeta)$, where $\alpha(\zeta)$ is directly correlated to the laser field. In this paper, we adopt normalized units, where length is scaled with the laser center wavelength ($\lambda_0$), velocity is scaled with the speed of light ($c$), momentum is scaled with $m_e c$, and electric (magnetic) field is scaled with $m_e \omega_0 c/e$ $(m_e \omega_0/e)$, respectively. Here, $m_e$ denotes the electron mass, $\omega_0$ denotes the laser center frequency, and $e$ represents the elementary charge.

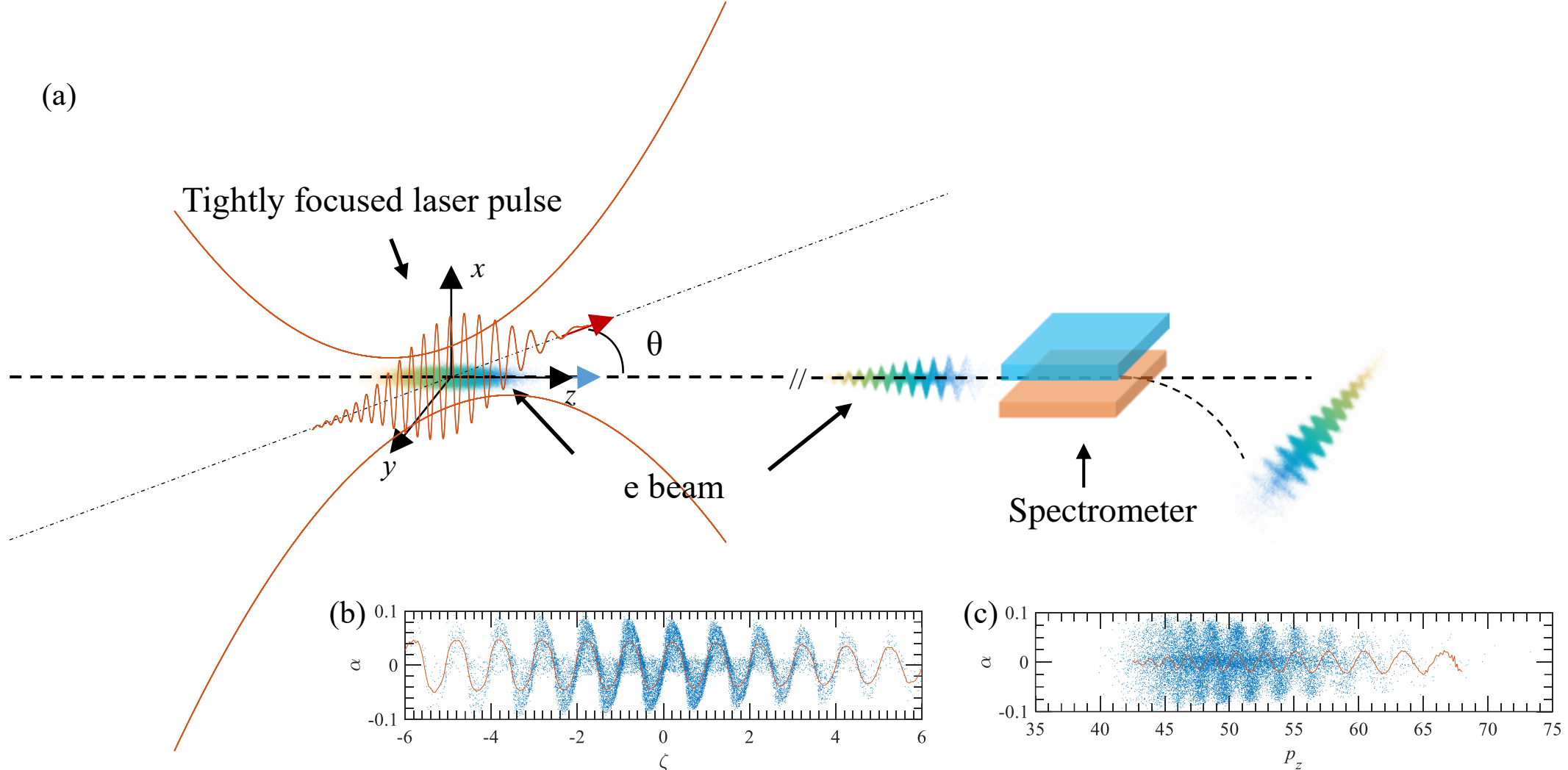

**Fig. 1.** (a) Schematic sketch of the e-beam chirp reconstruction via a tightly focused laser pulse. The e beam propagates along the $z$ axis, while the x-polarized laser pulse propagates in the y-z plane with a crossing angle of θ with respect to $z$. The interaction between the laser pulse and the e beam leads to longitudinal modulation of the e beam (b), and then the energy spectrum of the modulated e beam is detected by a spectrometer (c).

We first explore the process of momentum modulation in the e-beam during the interaction with the laser. Specifically, for the on-axis ($x$=$y$=0) electrons with $E_y = B_x = 0$, the momentum modulation at the laser polarization direction ($x$) can be described as $\Delta p_x = -e\int_{-\infty}^{\infty}(E_x - v_z B_y)dt$, the electrons are assumed to be relativistic with $v_z \approx 1$. In the case of a laser pulse propagating at a small crossing angle of $\theta$ with respect to the co-moving e beam, the on-axis ($z$) electric field, under the paraxial approximation, can be expressed as

$$E_x = a_0 e^{-\frac{r^2}{w_z^2}} \sin\left[-k\left(\zeta' + \frac{r^2}{2R_z}\right) + \psi_z + \psi_0\right], \quad (1)$$

where $w_z = w_0\sqrt{1+\left(\frac{z\cos\theta}{z_r}\right)^2}$, $r = -z\sin\theta$, $\frac{1}{R_z} = \frac{z\cos\theta}{(z\cos\theta)^2+z_r^2}$, $\psi_z = arctan\left(\frac{z\cos\theta}{z_r}\right)$, $\psi_0$ is the initial phase, $\zeta' = z\cos\theta - t$. $a_0$ is the normalized amplitude of the vector potential. Combining with $B_y = -\int\frac{\partial E_x}{\partial z}dt$, the transverse momentum gain $\Delta p_x$ of electron with initial retarded time of $\zeta_0$ could be derived as,

$$\Delta p_x = -e\int_{-\infty}^{\infty}(E_x - B_y)dt \approx \frac{2ea_0[2-(5\varepsilon^2+4)\cos^2\theta+(5\varepsilon^4+5\varepsilon^2+2)\cos^4\theta]\sin(k\zeta_0+\psi_0)}{5k\varepsilon^4\cos^4\theta}, \quad (2)$$

where $\varepsilon = \frac{w_0}{z_r}$ and $w_0$ is the laser waist, $z_r = \omega_0 w_0^2/2$ is the Rayleigh length. The optimal crossing angle, denoted as $\theta_m$, can be determined by finding the maximum value of $\Delta p_x$ while considering $\frac{d\Delta p_x}{d\theta} = 0$. Through calculations, we find that $\theta_m = \arctan\left(\pm\frac{\sqrt{5}}{2}\varepsilon\right)$. This derived value of $\theta_m$ coincides with the results obtained from particle simulations with a fifth-order correction description of fields, as depicted in Fig. 2 (a) (refer to the supplemental material for more details). It should be noted that within the range of $\tan\theta \in [0.3\varepsilon\ 2.1\varepsilon]$, $\Delta p_x$ exceeds $a_0$, indicating significant momentum modulation even when $\theta$ deviates significantly from $\theta_m$. Therefore, the introduction of a crossing angle $\theta$ provides a robust scheme for driving intense e-beam momentum modulation. Additionally, the temporal resolution of this scheme is directly related to the field period, with the potential for attosecond-scale temporal resolution achievable through the use of laser pulses with shorter wavelengths, such as femtosecond pulses generated through harmonic generation [35]. Although the use of shorter wavelength lasers may result in lower $a_0$ and weaker modulation, the transverse momentum modulation $\Delta p_x$ remains sufficiently large. Specifically, as shown in Fig. 2(b), $\Delta p_x$ can be enhanced to more than $4.5a_0$ at $\theta_m$, leading to a divergence amplitude modulation of approximately $45a_0$ mrad for electrons with $p_z$=100.

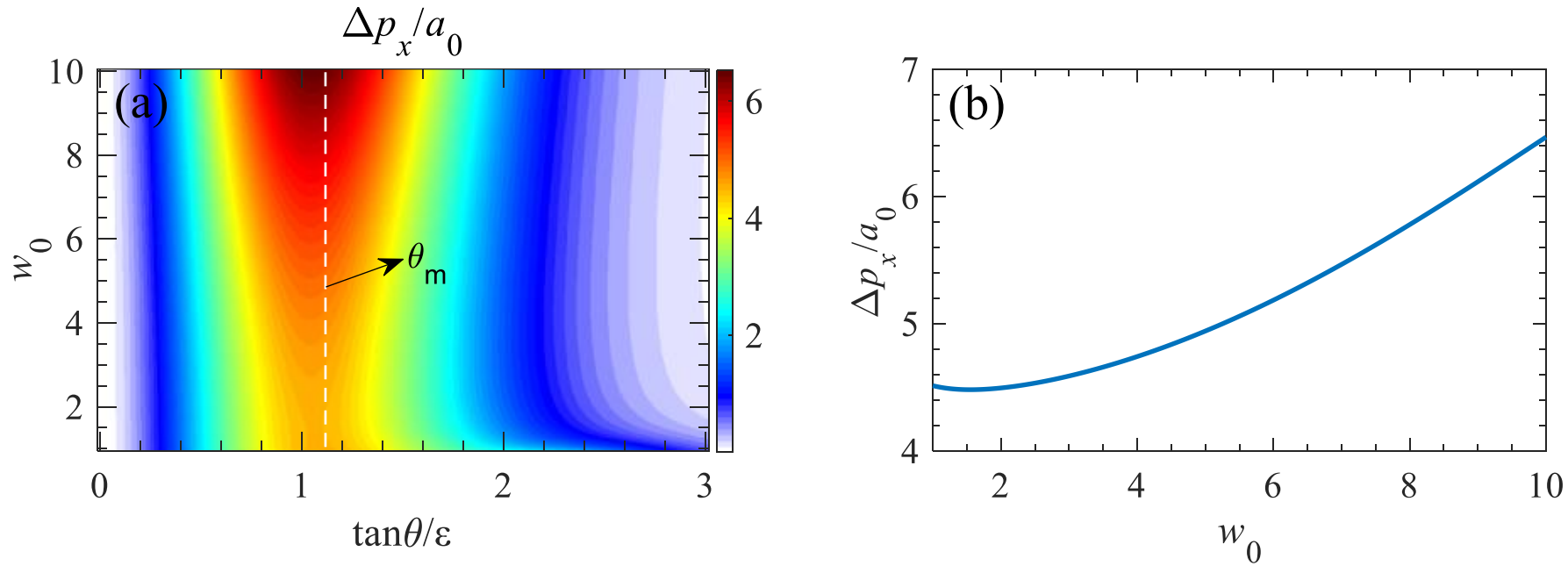

**Fig. 2.** (a) The momentum gain $\Delta p_x$ of on-axis electrons at $\zeta_0 = 1/4$ with $\psi_0 = 0$ and $p_z = 50$ for laser pulse with different $w_0$ while changing $\tan\theta$ from 0 to $3\varepsilon$, the $\theta_m = \arctan\left(\frac{\sqrt{5}}{2}\varepsilon\right)$ is marked. (b) The corresponding momentum gain $\Delta p_x$ at $\theta_m$. The fields description of fifth order correction is applied.

The modulation profile of $\Delta p_x(\zeta)$ is dependent on the witnessed fields. Figure 3 illustrates that the amplitude of $\Delta p_x$ for $\theta = \theta_m$ is significantly higher compared to the case of $\theta = 0$, attributed to the stronger resultant field $(E_x - B_y)$ and less phase mismatch. The modulation profile of $\Delta p_x(\zeta)$ follows the envelope of the laser pulse, expressed as $\Delta p_x(\zeta) \propto e^{-\zeta^2/\tau_L^2}$. Additionally, as depicted in Fig. 3(c), electrons separated transversely (both on and off-axis) exhibit the same modulation phase, and the modulation period aligns with the laser field along ζ, leading to $\Delta p_x(\zeta) \propto e^{i(\omega\zeta+\psi_0)}$. Therefore, the inferred modulation profile can be described as $\Delta p_x(\zeta) = C_p e^{-\zeta^2/\tau_L^2} e^{i(\omega\zeta+\psi_0)}$, where $C_p$ represents the modulation amplitude. In addition, to achieve transient reconstruction, it is essential to minimize the scale of the interaction time. As depicted in Fig. 3(d), $\Delta p_x$ rapidly increases as $N_r$ grows at the initial stage, reaching $2.3a_0$ within the interaction length from $-z_r$ to $z_r$. This indicates that a significant portion of transverse momentum is gained within a short interaction length $(-z_r,\ z_r)$. Consequently, a tightly focused laser pulse is necessary to reduce $z_r$ and obtain a shorter interaction length.

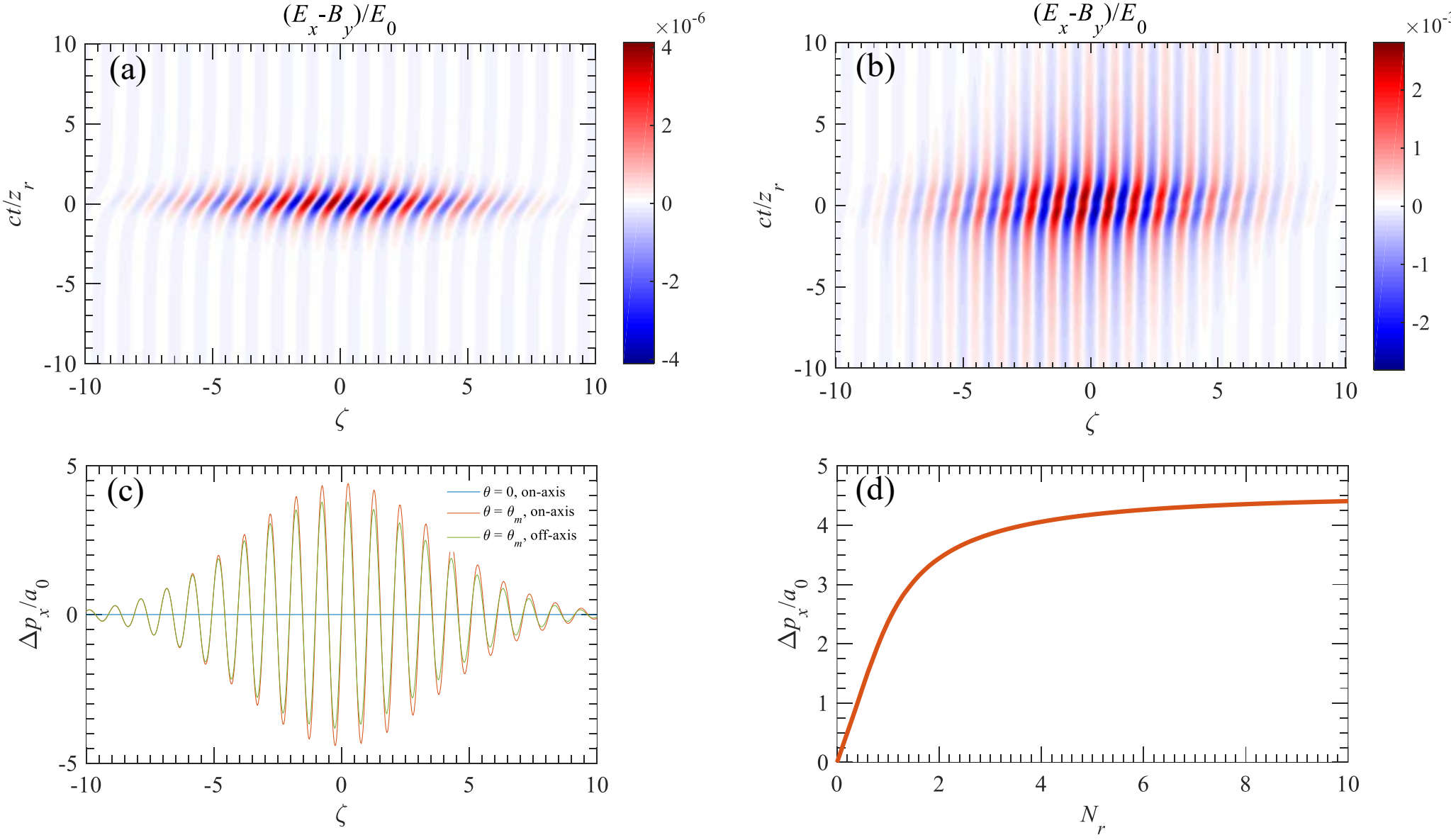

**Fig. 3.** The field $(E_x - B_y)$ the on-axis electrons witnessed during the laser-electrons interplay with $\theta = 0$ (a) and $\theta = \theta_m$ (b), (c) The $\Delta p_x$ integrated from $-10z_r$ to $10z_r$ for cases of (a), (b) and off–axis ($x = y = w_0$) electrons with $\theta = \theta_m$. (d) The maximum $\Delta p_x$ integrated from $-N_r z_r$ to $N_r z_r$ for case of (b). The other laser parameters: duration $\tau_L = 5,\ w_0 = 5$.

## 3. E-beam transient chirp reconstruction

Considering a linearly chirped laser with $\omega(\zeta) = \omega_0 + b\omega_0^2(\zeta + \tau_d)$, the interaction between the laser and the e-beam leads to longitudinal dispersion in the momentum modulation of the e beam, denoted as $p_x(\zeta)$. In our scheme, the intense modulation satisfying $p_z \gg \Delta p_x \gg p_{x0}$, where $p_{x0}$ represents the initial transverse momentum of the e-beam prior to the interaction, and $p_z$ is the longitudinal momentum of the e-beam, which remains almost unchanged during the interaction. Consequently, the e-beam divergence, denoted as $\alpha(\zeta) = p_x/p_z$, is correlated with $\omega(\zeta)$ and can be expressed as $\alpha(\zeta) = C_\alpha e^{-\zeta^2/\tau_L^2} e^{i(\omega\zeta+\psi_0)}$, where $C_\alpha = C_p/p_z$. To obtain a full temporal reconstruction of the e beam, the duration of the laser pulse $\tau_L$ should exceed $\tau_e + |\tau_d|$, where $\tau_e$ represents the duration of the e-beam and $\tau_d$ is the time delay between the centroid of the laser pulse and the center of the e-beam. Assuming the retarded origin ($\zeta$=0) is located at the e-beam center, then

$$\alpha(\zeta) = \begin{cases} C_\alpha e^{i([\omega_0 + b\omega_0^2(\zeta+\tau_d)](\zeta+\tau_d)+\psi_0)} e^{-(\zeta+\tau_d)^2/\tau_L^2}, & \zeta \in [-\tau_e, \tau_e] \\ 0, & \zeta \in \text{others} \end{cases}. \tag{3}$$

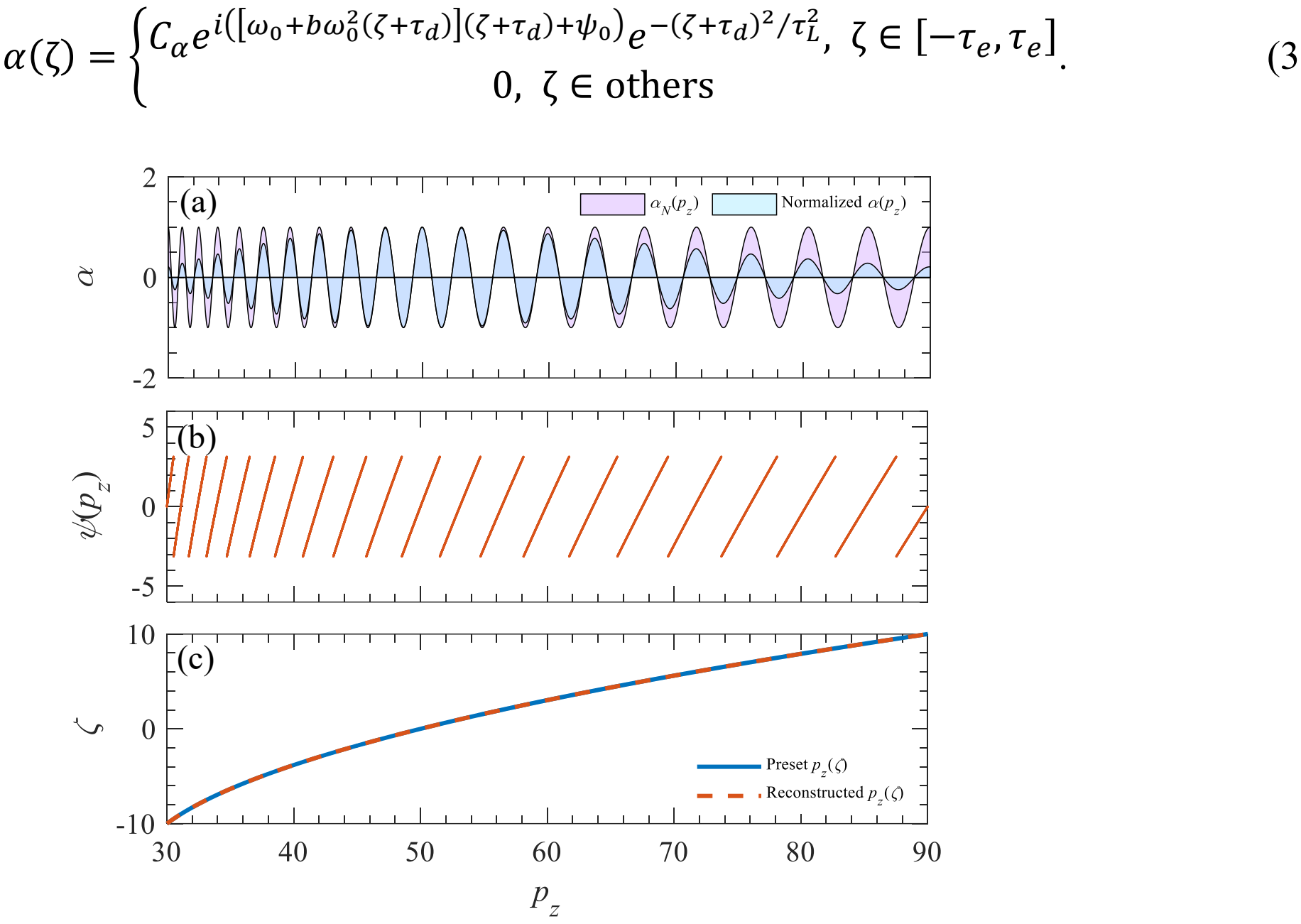

**Fig. 4.** E-beam chirp reconstruction from the divergence modulation. (a) The normalized e-beam divergence modulation $\alpha(p_z)$ attainable from a spectrometer and the corresponding $\alpha_N(p_z)$ normalized to its envelop. (b) The reconstructed phase profile $\psi(p_z)$ and (c) the reconstructed $p_z(\zeta)$ correlation. The e-beam parameters were set as $p_z = \sum_{i=0}^{2} C_i \zeta^i$ with $C_0 = 50$, $C_1 = 3$, $C_2 = 0.1$, and the modulation function $\alpha(\zeta) = e^{-\zeta^2/64} e^{2\pi i\zeta}$ was pre-defined.

To reconstruct the $p_z - \zeta$ correlation, the phase relation between the modulated divergence $\alpha(p_z)$ and $\alpha(\zeta)$ can be utilized, where $\alpha(p_z)$ can be obtained from an e-beam spectrometer. As shown in Fig. 4, in order to obtain the phase $\psi(p_z)$ of $\alpha$, $\alpha(p_z)$ should be normalized to its envelope $\alpha_N(p_z) = \alpha(p_z)/envelope[\alpha(p_z)]$, then an inverse cosine operation should be applied as $\psi_c(p_z) = \arccos[\alpha_N(p_z)]$, and the phase in one modulation period can be reconstructed as

$$\psi(p_z) = -i \ln[\cos\psi_c(p_z) + i\ \sin\psi_c(p_z)]. \tag{4}$$

As depicted in Fig. 4(b), the reconstructed phase $\psi(p_z)$ is confined to the range of [-π, π] within each period. By considering the phase correlation $\psi(p_z(\zeta)) = \psi(\zeta) = \omega(\zeta)(\zeta+\tau_d)+\psi_0$, the complete $p_z - \zeta$ correlation can be obtained by connecting these individual profiles, as shown in Fig. 4(c), the reconstructed $p_z - \zeta$ correlation coincides with the pre-set $p_z(\zeta)$ profile.

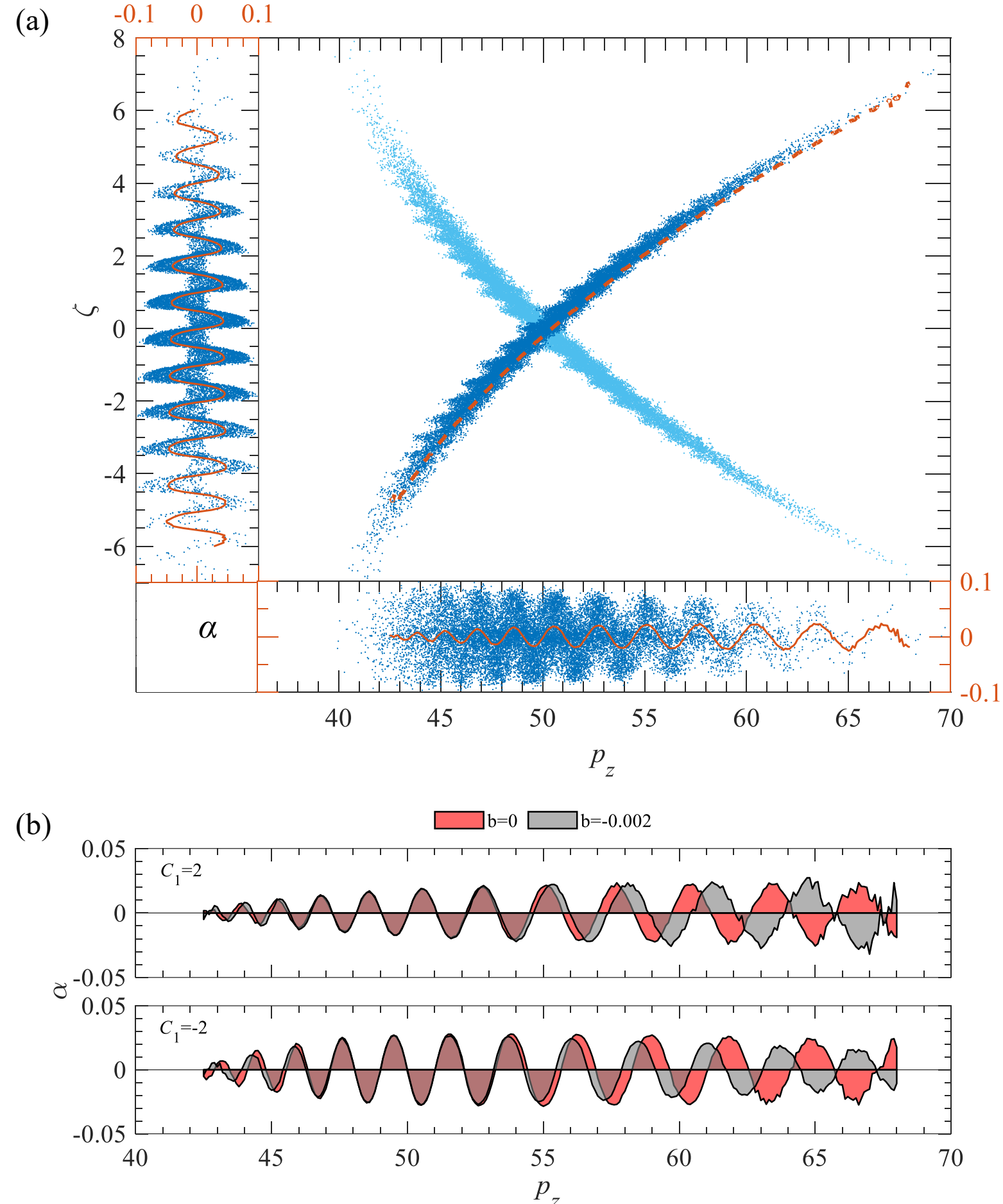

**Fig. 5.** The schematic of the e-beam chirp reconstruction. (a) A typical e-beam phase space distribution after the interplay with unchirped laser pulse ($b$=0), the inner insets show the corresponding modulation projection $\alpha(\zeta)$ and $\alpha(p_z)$, the dashed red line is the reconstructed e-beam chirp profile $\psi(p_z)/\omega_0$. (b) $\alpha(p_z)$ for cases of negatively ($C_1 = 2$) and positively ($C_1 = -2$) chirped e beams with unchirped ($b$=0 (red)) and chirped ($b$=-0.002 (grey)) laser pulses applied.

Particle simulations were carried out to validate the feasibility of the proposed scheme, employing the fifth-order correction description of fields. In the simulations, the laser pulse parameters were set as follows: $w_0 = 5$, $\tau_L = 5$, $\theta = \theta_m$, and $a_0 = 1$, and the Gaussian e-beam parameters were set to be: $\tau_e = 2$, transverse emittance of 2 (corresponding to $2\lambda_0$ m $\cdot$ rad in the SI units) and rms space spread of $\sigma_x = \sigma_y = 2$, the e-beam chirp profile is set to be $p_z = \sum_{i=0}^{2} C_i \zeta^i$ with $C_0 = 50$ and $C_2 = 0.2$. Figure 5(a) demonstrates the correlation between $p_z$ and $\zeta$ for the case of an unchirped laser pulse (b=0) with $\psi\big(p_z(\zeta)\big) = \psi(\zeta) = \omega_0(\zeta + \tau_d) + \psi_0$, By shifting the phase of the e-beam center to 0, the reconstructed $\zeta = \psi(p_z)/\omega_0$ agrees with the designed e-beam phase space distribution. However, since the relative longitudinal position information of individual electrons is not available from a spectrometer, the reconstructed e-beam chirp for a monotonous case can be positive or negative, as depicted in Fig. 5(a). Therefore, in our scheme, a chirped laser pulse with b≠0 is further required to resolve the temporal correlation of the e-beam energy. Figure 5(b) illustrates the effect of a non-zero chirp parameter (b) on the e-beam chirp. When a positively chirped pulse with b<0 is applied, if the modulation period is longer (shorter) at the higher energy region, the e-beam exhibits a negative (positive) chirp.

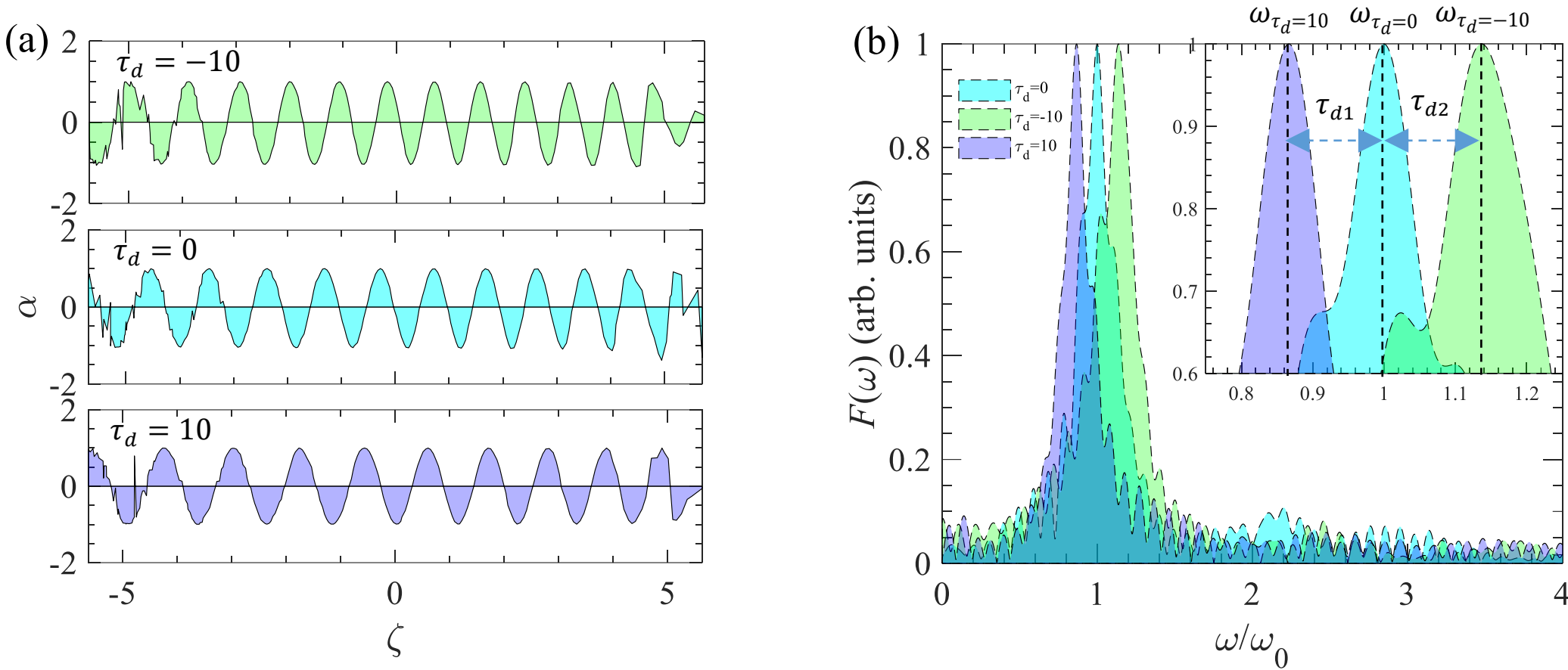

**Fig. 6.** The derived divergence modulation $\alpha(\zeta)$ (a) and the corresponding $F(\omega)$ (b) for various $\tau_d$ ($\tau_d = -10$ (green), $\tau_d = 0$ (cyan), $\tau_d = 10$ (blue) ) based on the reconstructed $p_z(\zeta)$, employing a chirped laser pulse with b=-0.002.

Additionally, the utilization of a chirped laser pulse enables the estimation of the delay $\tau_d$ between the e-beam and the center of the laser pulse, making it suitable for various advanced applications such as timing jitter probe in UED [10, 11] and temporal control of multiple beams in PAs [41, 42]. In the frequency domain, the locally modulated frequency of the e-beam center can be expressed as $\omega = \omega_0 + b\omega_0^2\tau_d$, allowing the estimation of $\tau_d$ through the frequency deviation $\tau_d = (\omega - \omega_0)/b\omega_0^2$. To obtain real-time $\omega$, two steps are performed. Firstly, the divergence modulation $\alpha(\zeta)$ is derived based on the reconstructed $p_z(\zeta)$, as illustrated in Fig. 6(a). Secondly, based on the derived $\alpha(\zeta)$, the shifted $\omega$ can be obtained through the Fourier transform of $\alpha(\zeta)$ as $F(\omega) = \int_{-\infty}^{\infty} \alpha(\zeta) e^{-i\omega\zeta} d\zeta$, as shown in Fig. 6(b). For the three simulated cases with $\tau_d = \pm 10$ and 0, the corresponding reconstructed peak frequencies were $\omega_{\tau_d=10} = 0.870\omega_0$, $\omega_{\tau_d=-10} = 1.131\omega_0$, $\omega_{\tau_d=0} = 1.000\omega_0$. Consequently, the delay can be estimated as $\tau_{d1} = (\omega_{\tau_d=10} - \omega_{\tau_d=0})/b\omega_0^2 = 10.3$ and $\tau_{d2} = (\omega_{\tau_d=-10} - \omega_{\tau_d=0})/b\omega_0^2 = -10.4$, which are consistent with the designed parameters.

Efficient discrimination of the modulation migration of $\alpha(p_z)$ along $p_z$ in a spectrometer is a crucial aspect of this scheme. It involves two types of migration: adjacent period migration ($\Delta p_{m1}$) and modulation migration ($\Delta p_{m2}$) between the unchirped and chirped laser cases. For the former with $b$=0, since $\Delta p_z = \sum_{i=1}^{n} iC_i\zeta^{i-1}\Delta\zeta$, the migration for an increment of $\Delta\zeta$=1 is given by $\Delta p_{m1} = \left|\sum_{i=1}^{n} iC_i\zeta^{i-1}\right|$. As for the latter, considering the frequency-space relation $\Delta\omega = -2\pi\Delta\zeta/\zeta^2$ and the frequency discrepancy between the unchirped and chirped laser pulse ($\Delta\omega = b\omega_0^2(\zeta+\tau_d)$), the corresponding migration is $\Delta p_{m2} = \left|\frac{b\omega_0^2(\zeta+\tau_d)}{2\pi}\sum_{i=1}^{n} iC_i\zeta^{i+1}\right|$. To ensure the feasibility of the scheme, the resolution of the e-beam spectrometer $\delta_p$ at a local $\zeta$ or the corresponding $p_z(\zeta)$ should be smaller than the minimum value between $\Delta p_{m1}$ and $\Delta p_{m2}$, expressed as $\delta_p < min(\Delta p_{m1}, \Delta p_{m2})$.

## 4. Conclusion

In conclusion, we have presented a novel method for measuring the transient energy chirp of an ultra-short e-beam using a tightly focused laser pulse. Through detailed exploration of the laser-electron interaction, we have shown that a crossing angle $\theta_m$, approximately given by $\tan\theta_m = \pm\frac{\sqrt{5}}{2}\varepsilon$, can significantly enhance the modulation effect. By leveraging the phase relation between the modulated divergence $\alpha(p_z)$ and $\alpha(\zeta)$, we can provide real-time and precise characterization of the e-beam transient phase space. Furthermore, the delay between the e-beam and laser pulse can be directly estimated using the Fourier transform of the reconstructed $\alpha(\zeta)$. With the utilization of high-efficiency laser harmonic techniques, the resolution of this versatile method can be extended to the attosecond regime. This facilitates the optimization and stabilization of PAs, enables attosecond UED timing jitter probes, and advances the field of attosecond sources and attosecond structural dynamics.

**Funding.** This work was supported by the National Natural Science Foundation of China (Grants Nos. 11905279, 12005137), the Innovation Program of Shanghai Municipal Education Commission (Grant No. 2021-01-07-00-02-E00118), and sponsored by Shanghai Sailing Program (Grants No. 20YF1435400).

**Disclosures.** The authors declare no conflicts of interest.

**Data availability.** Data underlying the results are not publicly available at this time but may be obtained from the authors upon reasonable request.

**Supplemental document.**

# Supplementary Material

## I. High order correction description of the laser fields

The fifth order correction description of the laser pulse field components with an angle of $\theta$ rotating about the x-axis are [43]

$$E_x = -iE\left\{1 + \varepsilon^2\left(f^2\xi^2 - \frac{f^3\rho^4}{4}\right) + \varepsilon^4\left[\frac{f^2}{8} - \frac{f^3\rho^2}{4} - \frac{f^4}{16}(\rho^4 - 16\rho^2\xi^2)\right.\right.$$

$$\left.\left. - \frac{f^5}{8}(\rho^6 + 2\rho^4\xi^2) + \frac{f^6\rho^8}{32}\right]\right\}, \tag{S1}$$

$$E_y = E_y' \cos\theta - E_z' \sin\theta, \tag{S2}$$

$$E_z = E_y' \sin\theta + E_z' \cos\theta, \tag{S3}$$

$$B_x = 0, \tag{S4}$$

$$B_y = B_y' \cos\theta - B_z' \sin\theta, \tag{S5}$$

$$B_z = B_y' \sin\theta + B_z' \cos\theta, \tag{S6}$$

with

$$E_y' = -iE\xi\upsilon\left[\varepsilon^2 f^2 + \varepsilon^4\left(f^4\rho^2 - \frac{f^5\rho^4}{4}\right)\right], \tag{S7}$$

$$E_z' = E\xi\left[\varepsilon f + \varepsilon^3\left(-\frac{f^2}{2} + f^3\rho^2 - \frac{f^4\rho^4}{4}\right) + \varepsilon^5\left(-\frac{3f^3}{8} - \frac{3f^4\rho^2}{8} + \frac{17f^5\rho^4}{16} - \frac{3f^6\rho^6}{8} + \frac{f^7\rho^8}{32}\right)\right], \tag{S8}$$

$$B_y' = -iE\left\{1 + \varepsilon^2\left(\frac{f^2\rho^2}{2} - \frac{f^3\rho^4}{4}\right) + \varepsilon^4\left[-\frac{f^2}{8} + \frac{f^3\rho^2}{4} + \frac{5f^4\rho^4}{16} - \frac{f^5\rho^6}{8} + \frac{f^6\rho^8}{32}\right]\right\}, \tag{S9}$$

$$B_z' = E\upsilon\left[\varepsilon f + \varepsilon^3\left(\frac{f^2}{2} + \frac{f^3\rho^2}{2} - \frac{f^4\rho^4}{4}\right) + \varepsilon^5\left(\frac{3f^3}{8} + \frac{3f^4\rho^2}{8} + \frac{3f^5\rho^4}{16} - \frac{4f^6\rho^6}{4} + \frac{f^7\rho^8}{32}\right)\right], \tag{S10}$$

where $\xi = x/w_0$, $\upsilon = (y\cos\theta - z\sin\theta)/w_0$, $\varepsilon = w_0/z_r$, $\rho^2 = \xi^2 + \upsilon^2$, $f = i/[(y\sin\theta + z\cos\theta)/z_r + i]$, $z_r = \omega_0 w_0^2/2$ is the Rayleigh length, $w_0$ is laser waist size, $E = a_0 f \exp(-f\rho^2)\exp\{-i[\omega(\zeta)\zeta + \varphi_0] - \zeta^2/\tau^2\}$, $\omega(\zeta) = \omega_0 + b\omega_0^2\zeta$ for a linearly chirped laser pulse, $\zeta = y\sin\theta + z\cos\theta - t$ is the retarded time, *b* is the frequency chirp coefficient, $\omega_0 = 2\pi$ is the frequency at $\zeta = 0$, $\tau$ is the laser pulse duration, $\varphi_0$ is the constant phase.